\documentclass[10pt,reprint]{socc14}

\PassOptionsToPackage{hyphens}{url}\usepackage[colorlinks=true,urlcolor=blue,citecolor=blue,linkcolor=blue,plainpages=false]{hyperref}
\usepackage[normalem, normalbf]{ulem} 

\usepackage{times}
\usepackage{listings, textcomp}
\usepackage[T1]{fontenc}
\usepackage{xcolor}
\usepackage{multirow}
\usepackage{booktabs}
\usepackage{graphicx}
\usepackage{enumitem}
\usepackage{changes}
\usepackage{amsmath}
\usepackage{mathptmx}
\usepackage[absolute]{textpos}
\usepackage[TABBOTCAP]{subfigure}

\makeatletter
\def\@copyrightspace{\relax}
\makeatother

\begin{document}

\setlength{\pdfpageheight}{\paperheight}
\setlength{\pdfpagewidth}{\paperwidth}

\conferenceinfo{Submission to SoCC '15}{April, 2015, Kohala, Hawaii, USA} 
\copyrightyear{2015} 
\copyrightdata{978-1-nnnn-nnnn-n/yy/mm} 
\doi{nnnnnnn.nnnnnnn}


\exclusivelicense                



\title{\bf Distributed Graphical Simulation in the Cloud}

\authorinfo{Omid Mashayekhi\and Chinmayee Shah\and Hang Qu\and Andrew Lim\and Philip Levis}
           {Stanford University}
           {\{omidm,~chshah,~quhang,~alim16\}@stanford.edu \and pal@cs.stanford.edu}


\maketitle

\setlength{\TPHorizModule}{1mm}
\setlength{\TPVertModule}{1mm}
\begin{textblock}{200}(110,87)
\end{textblock}

\begin{abstract}
Graphical simulations are a cornerstone of modern media and films.
But existing software packages are designed to run on HPC nodes,
and perform poorly in the computing cloud.
These simulations have complex data access patterns over complex data
structures, and mutate data arbitrarily, and so are a poor fit for existing
cloud computing systems.
We describe a software architecture for running graphical simulations
in the cloud that decouples control logic, computations and data exchanges.
This allows a central controller to balance load by redistributing
computations, and recover from failures.
Evaluations show that the architecture can run existing, state-of-the-art 
simulations in the presence of stragglers and failures, thereby
enabling this large class of applications to use the computing cloud for
the first time.




\end{abstract}



% ...

\section{Introduction}

Graphical simulation is a staple of modern digital entertainment.  When we see
a river flow in the movie Brave, an explosion in Star Wars: Revenge of the
Sith, or smoke billowing from destroyed buildings in Man of Steel, we see the
result of computationally simulating fluids: water, fire, and smoke.

Being able to run simulations in the cloud would enable studios to
elastically scale their simulation infrastructure when needed, such as
during final production, when each shot has its final render.
Graphical simulation software packages are designed to run on a single
powerful server or small, 3-4 node high performance computing clusters
with InfiniBand~\cite{infiniband} as their interconnect. The
techniques and algorithms these simulations use work poorly in the
cloud.  They assume that all nodes can communicate
equally, all nodes run at exactly the same speed, and failures are
very rare (e.g., $<1$ in 40,000 in a multi-day simulation). They
evenly partition the simulation across all of the cores used, so the
simulation runs as fast as the slowest core.
To handle rare failures, they use expensive and infrequent checkpointing
mechanisms. Furthermore, parallel nodes run in lockstep, such that the
high latency of Ethernet (100 microseconds, rather than 500
nanoseconds with InfiniBand) causes cores to fall idle during
communications.

Graphical simulations require very different data and execution models than
what current cloud computing systems provide. A graphical
simulation uses multiple complex data models, such as
a marker-and-cell grid~\cite{mac} for the fluid volume, a dense particle
field for the fluid surface~\cite{particle-levelset}, and a system
of linear equations to ensure fluid does not disappear. 
These data structures are geometric in nature and computations on neighboring
regions have tight dependencies.
A simulation involves a loop of many iterations that advance time. All 
simulation state is held in memory, as I/O is far too slow.  These
requirements differ greatly from data tuples as in MapReduce~\cite{mapreduce},
Spark~\cite{spark}, and Naiad~\cite{naiad} or graphs as in
Pregel~\cite{pregel} and PowerGraph~\cite{powergraph}.

This paper presents Nimbus, a system for running graphical simulations
in the computing cloud. To deal with the scheduling challenges inherent
to cloud systems, Nimbus, like other cloud systems, uses a centralized
{\it controller} node that is responsible for monitoring the entire
state of the simulation. To enable dynamic load balancing, Nimbus
decouples data exchange and the simulation execution plan. The system
runtime is responsible for data exchanges between nodes, and
invoking a simulation function after all its data is ready. This
decoupling gives Nimbus the ability to place data and computation based
on global knowledge of the system. To make applications tolerant
to node failures, the controller continuously monitors progress and 
dynamically inserts check-points to save data, as needed.



The next section provides an overview of graphical simulations,
which motivates a set of requirements for a system to support
them in the cloud.  Section~\ref{sec:design} presents a system design
whose abstractions meet these requirements.
Section~\ref{sec:implementation} details implementation, and.
Section~\ref{sec:evaluation} evaluates how the
system handles stragglers and node failures. Section~\ref{sec:related} and
Section~\ref{sec:future} conclude with related work and a set of open questions
for future work.

%
%
%
%
%
%

\section{Graphical Simulations} \label{sec:motivation}

Graphical simulations use different data models and algorithms 
than what available cloud frameworks provide.
This section gives an overview of the principal methods and algorithms used in
graphical simulations, and explains the challenges of distributing these
computations over multiple nodes. The nature of these simulations and the
associated challenges motivate a set of system design
requirements~(\S\ref{sec:design}).

As a concrete example of a graphical simulation, we focus on
PhysBAM~\cite{physbam}, an open source physics based software package
for fluid and rigid body simulations. Movie studios such as ILM and
Pixar use PhysBAM in production films, and the
developers have won two Academy Awards for its contributions to
special effects~\cite{oscars}. PhysBAM can simulate a huge
range of phenomena, but in the rest of this paper, we focus on a water simulation.
Water simulation is a canonical example, as it is extremely difficult and employs methods
that are required for other fluid simulations such as smoke and fire.

\subsection{Fluid Models and Simulation Algorithms} 
\label{sec:particle-levelst}

There are two basic ways to computationally represent a fluid: a grid
or particles. A grid divides simulated volume into cells. Per-cell
state describes the state of the simulation, such as whether it
contains fluid, pressure, and velocity.  The second approach is to
represent the fluid as a set of particles, each of which has its own
$(x,y,z)$ coordinates, velocity, and size. Grids and particles have
different strengths and weaknesses. For example, a grid smoothes out
small ripples but do not model splashes well, while particles have
difficulty representing fixed boundaries such as the edge of a glass.

The particle-levelset
method~\cite{particle-levelset}, pioneered by
PhysBAM, combines particle and grid representations and is why movie
and special effect studios can simulate water, smoke, and fire today.
The key insight is that the most important visual feature is the
surface of the fluid.  The particle level-set method use a coarse
grid, augmented with dense particles \textit{only on the surface}.
Combining these two methods, however makes
simulations much more complex, as the grids and particles interact in
subtle and interesting ways.\footnote{For example, particles that
  leave the surface become drops in a splash, and must be correctly
  merged back with the water mass when they hit the surface again.
  Readers interested in a more complete description of the
  complexities can read the seminal book on the topic by Bridson~\cite{bridson}.}

A simulation is a loop: each iteration steps time 
forward.~\footnote{The length
of the time step is determined by fluid velocity and grid resolution, so
that fluid does not seem to leap through space.} When time passes
a frame boundary, the simulation outputs the visual state of the simulation
for later rending. An iteration
has 22 distinct computational steps, which can be divided into three
major categories: updating grid cells, updating particles, and solving
a set of linear equations that enforce physical laws on the water (e.g.,
it does not compress or disappear). Solving the linear equations uses
a sub-loop within the main loop. In a typical $256^3$ water simulation,
there are on average 20 main loop iterations per frame
(24fps means 42ms/frame, the main loop time step is 1.6ms) and 
100 iterations of the inner solver loop. Table~\ref{tab:sim-scale} 
shows where a time step spends its time.

\subsection{Current Distributed Simulations} \label{sec:distributed}

\begin{figure}
\centering
\includegraphics[width=3.5in]{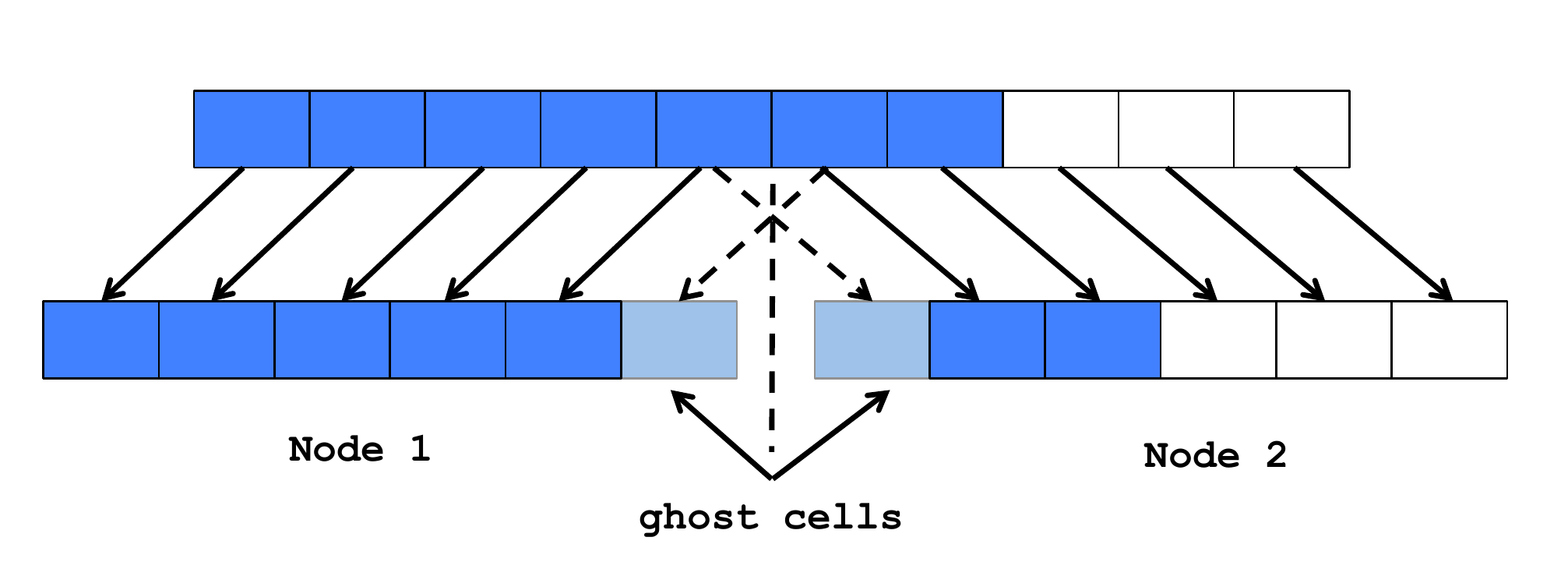}
\caption{A 1D row of water represented in a grid. When partitioned
across two processes, the two processes must exchange \textit{ghost cells} of
state so they can perform computations locally.}
\label{fig:row}
\end{figure}

Running a simulation across multiple nodes requires partitioning the
simulation geometry across them. The basic challenge is that partitions
are not independent. The state of water at any cell is dependent on
its neighboring cells, some of which may be on a different node.  
Furthermore, solving the linear equations involve global operations.

\begin{figure}
\includegraphics[width=3in]{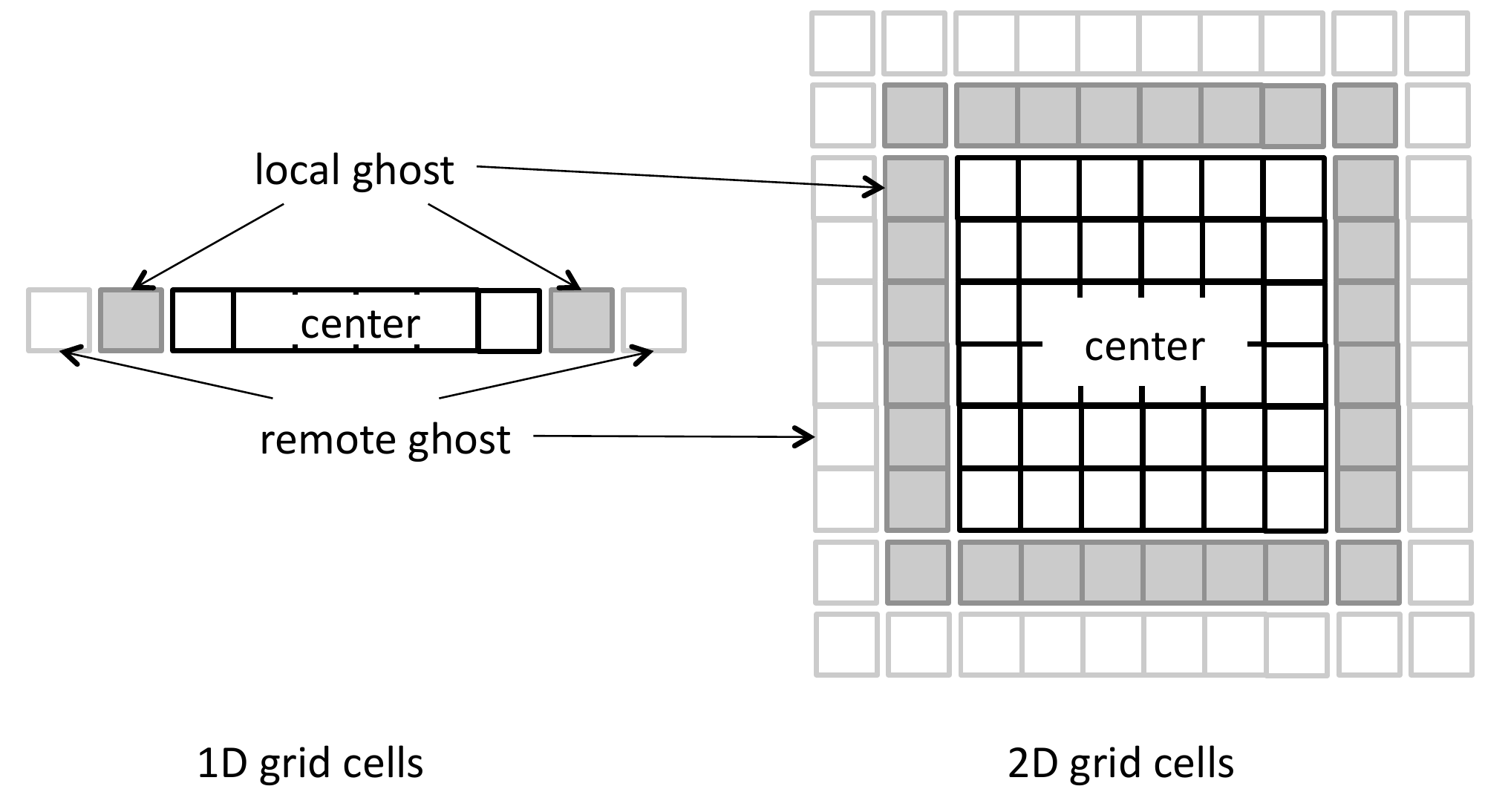}
\caption{Ghost cell configurations in simulation grids.
The local state on a node consists of $3^{d}$ objects, where
$d$ is the dimensionality of the grid, while the combined
local and remote state a node must use consists of $5^{d}$
objects. A 3D grid is not shown for visual simplicity: per-node
state is 27 objects and the total state is 125 objects.}
\label{fig:cells}
\end{figure}

Partitions can be distributed while minimizing data sharing with
\textit{ghost cells}. Consider a simple 1D
simulation of a pipe with water, shown in Figure~\ref{fig:row}.  Each
partition is divided into five parts per axis: a large,
central region that only the local computations need, two thin regions of ghost
cells that are sent to neighbors, and two thin regions of ghost cells
that are received from neighbors.  
Figure~\ref{fig:cells} shows a partition in a 1D and a 2D grid.  
For a 3D simulation, a partition consists of 125 separate regions ($5^{3}$).
Each variable is partitioned in this manner, resulting in over $29$ thousand
data objects for 16 partitions, in a typical simulation with 21 different
variables.


%
%
%
%

\begin{table}
\centering
\begin{tabular}{l r r r}
\toprule[1pt]
                      & one solver & particle- & entire    \\  
                      & iteration  & levelset  & main-loop \\  
\midrule[0.5pt]
computation substeps  & 4          & 22        & 422       \\
global reductions     & 2          & 2         & 202       \\
ghost value updates   & 1520       & 73.6K     & 225.6K    \\
duration              & 61ms       & 6.7s      & 12.91s    \\
\end{tabular}
\caption{Number of computation substeps, global reductions, and ghost value
  updates for a $256^3$ water simulation with 16 partitions.
  Each main-loop consists of particle-levelset
  operations in addition to $100$ solver iterations.}
\label{tab:sim-scale}
\end{table}

In addition to computing on particles and grid cells, simulations also need
to perform global reductions. For example, to compute the time step, or
the residual of the linear solve, the simulation takes the maximum value
across all of the partitions. Table~\ref{tab:sim-scale} summarizes
the number of computation substeps, global reductions, and ghost value updates
in the main-loop and its components for water simulation.

When a computational step (e.g., reseeding particles) completes, that
node needs to send the updates it made to local ghost regions and 
receive updates for remote ghost regions. PhysBAM does this in lockstep:
each worker process completes its computation, sends its results, then
blocks on receiving results from neighbors. This approach tightly
couples the control flow of the program with its state exchange. Furthermore,
the partitions are set up statically at compile time and cannot move.
If one node fails, the entire simulation fails. The simulation can
run only as fast as the slowest node in the cluster, so stragglers
are a major concern.

\subsection{Design Requirements}

Finally, while interacting with the PhysBAM developers and other
graphical simulation researchers, we learned that there is a strong
hesitation in changing available libraries and code bases. Core
libraries has been tested for correctness and optimized for
performance over many years. For example, PhysBAM library is over 50
developer-years worth of work and supports tens of applications.

In order to run these simulations in the cloud, in presence of stragglers and
failures, we derived the following three system requirements:

\begin{enumerate}[nolistsep]
\item the system's abstractions must allow dynamic data placement and load 
distribution for graphical simulations,
\item the runtime must schedule around stragglers and recover from failures, and
\item the system must be able to run existing simulation codes with minimal changes.
\end{enumerate}

Achieving the first two goals will enable simulations to run in the
cloud; achieving the third will mean there are simulations to run and
this capability will be an attractive option for developers.

\section{System Design} \label{sec:design}

This section presents a system design that addresses the
requirements listed in the previous section.
To satisfy the first requirement, we decouple control flow, computations and
data exchange.
Specifically, an application is decomposed into a series of \textit{jobs} with
pure computation and no communication. Each job has compact meta data that
determine data dependencies and job execution order.
Nimbus runtime deciphers and performs data exchanges between nodes
(for ghost values and reductions) based on this metadata, as required.

To address the second requirement, Nimbus uses a centralized controller that
maintains global information about performance of nodes, to detect stragglers
and failures. This is similar to other cloud computing
systems~\cite{mapreduce,spanner,spark}. The controller makes decisions about
data and job placement, load-balancing the simulation as stragglers appear. It
creates periodic checkpoints of the simulation state, and rewinds back, when one or more nodes fail.

Nimbus does not make any assumptions about data access and computation patterns
within computation jobs, except that computation jobs do not perform any data
exchanges on their own.
This allows us to use code from existing simulation libraries with
minimal changes and some additional code to specify metadata.
This helps us meet the third goal. This is covered in more detail in
\S~\ref{sec:porting}.

\subsection{Application Abstraction} \label{sec:abstraction}

Each variable over a simulation domain is decomposed into disjoint
\textit{data} objects over the ghost and central regions, as
depicted in Figure~\ref{fig:cells}.
Application logic is decomposed into a series of computation units, called
\textit{jobs}. Each job is characterized by four things: 
(i)~\textit{Read} set of data objects to read,
(ii)~\textit{Write} set of data objects to write,
(iii)~\textit{Before} set of jobs that must finish before the job starts
executing, and
(iv)~Computation code to perform the actual computation.
Before sets determine the control flow of the application, while read/write
sets determine the data requirements of a job. Before sets and read/write sets
comprise a job's metadata.

Jobs mutate data and/or spawn new jobs.
An application starts with a special job \verb+main+, which spawns new jobs.
Applications iterate by spawning jobs that spawn a batch of jobs.
When a job running on a node spawns a new job, the node submits the job
to a centralized controller for execution.
The controller assigns these spawned jobs to nodes, which execute the
corresponding simulation code.

\begin{figure}[t]
\centering
\begin{tabular}{c}
\subfigure[Simulation data as disjoint data.]
{
\includegraphics[width=3in]{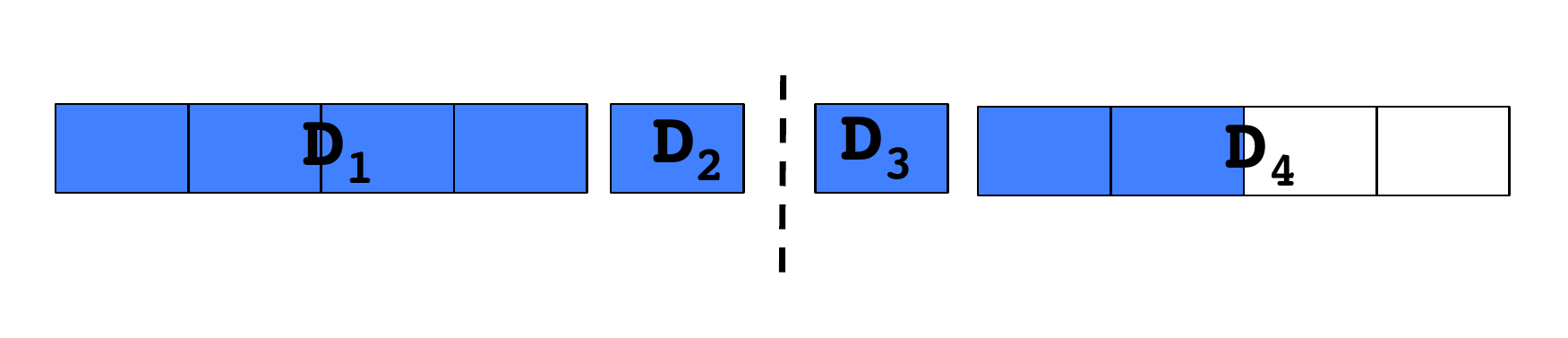}
\label{fig:example-data}
}
\\
\subfigure[Job graph for one simulation iteration, with each job's
meta-data. Each substep has two jobs that operate over left and right
partitions. Dashed jobs spawn new jobs.]
{
\includegraphics[width=3in]{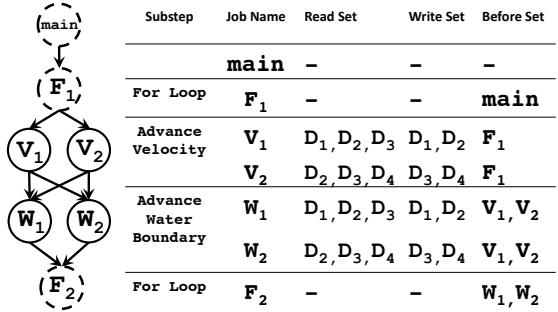}
\label{fig:example-jobs}
}
\end{tabular}
\caption{Discretized 1D water simulation example under Nimbus
  abstraction split into two partitions.}
\label{fig:abstraction}
\end{figure}

Figure~\ref{fig:abstraction} illustrates this with a simplified 1D water
simulation example over two partitions.
The simulation updates velocity for each cell that contains water, and then
moves water using the updated velocity.
The application comprises of four jobs -- \verb+main+ spawns
\verb+Forloop+, and \verb+ForLoop+ spawns \verb+AdvanceVelocity+,
\verb+AdvanceWater+ and conditionally, a new \verb+ForLoop+ job for the next
iteration.
Figure~\ref{fig:example-data} shows how velocity is decomposed into disjoint
central and ghost data objects.
Figure~\ref{fig:example-jobs} shows the metadata for each job.
Note that data objects over ghost
regions appear in read set of multiple jobs, while central data objects are
read/written by only one job, in each substep.
The \textit{job graph} in Figure~\ref{fig:example-jobs} depicts the application
flow.
Jobs with a dashed outline spawn new jobs for the next iteration.

Exchanging job metadata between controller and nodes quickly and storing
them in optimized data structures for fast queries is critical to runtime
performance. Data objects and jobs are represented using integer
identifiers, \textit{data id} and \textit{job id}.
This allows Nimbus to compactly represent metadata as integer sets
\footnote{A serialization implementation based on
protocol~buffer~\cite{protobuf} shows about 90\% compression ratio compared to
ASCI identifiers.}, and deploy efficient hash table for queries.

\subsection{Centralized Controller} \label{sec:centralized-controller}

\begin{figure}[t]
\centering
\includegraphics[width=3.5in]{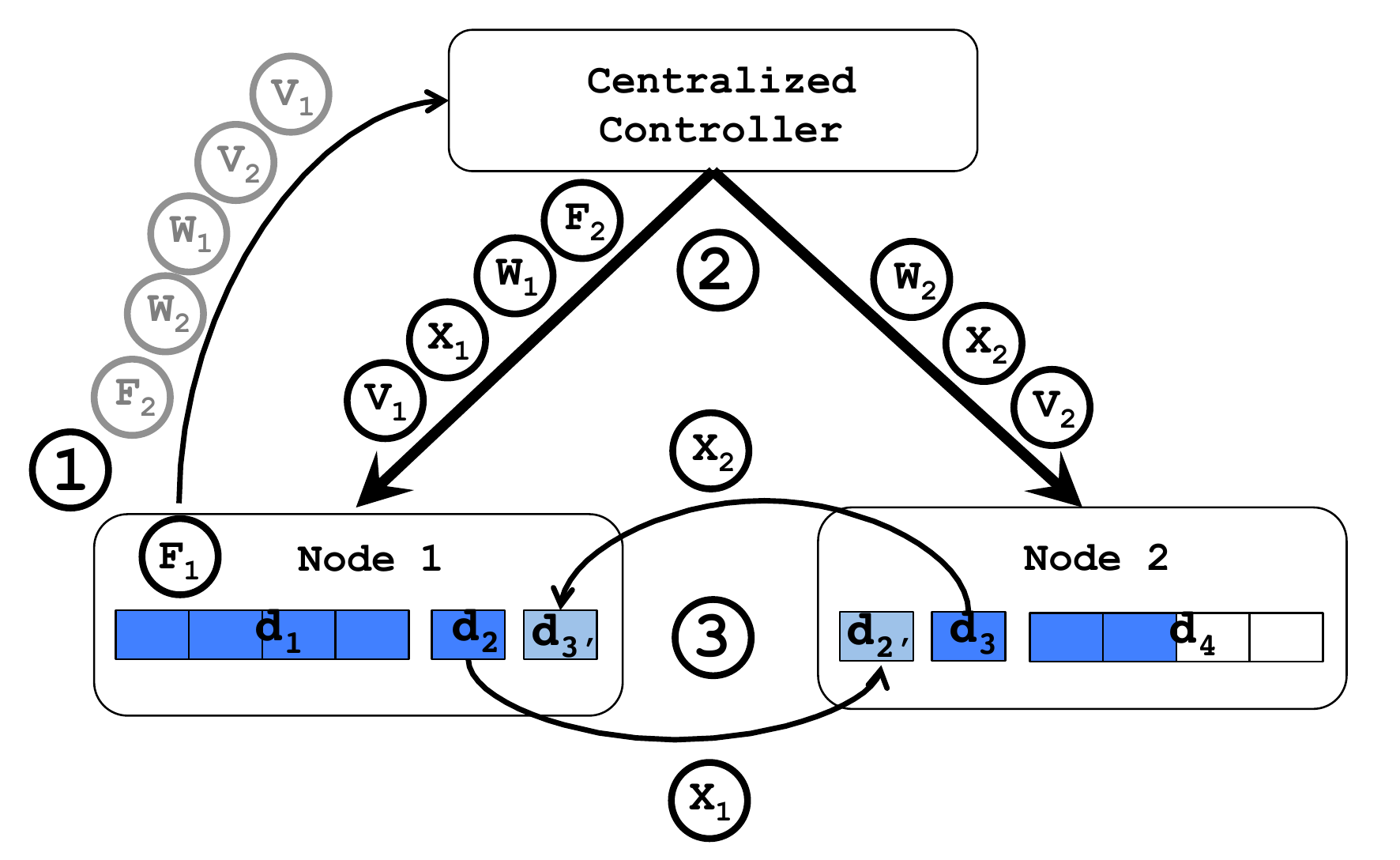}
\caption{Centralized controller driving 2 nodes for one iteration of
  discretized 1D  water simulation example. Nodes submit jobs to
    scontroller in phase~$1$. Controller instantiates data objects,
    inserts required data exchanges (X) in between compute jobs and
    assigns jobs to nodes  in phase~$2$. Computation and data
    exchange happen at the nodes in phase~$3$.}
\label{fig:architecture}
\end{figure}

Centralized controller monitors resources in the cloud and drives a
simulation over available resources by issuing commands to the nodes.
These commands instantiate data objects, assign computation jobs to nodes,
and exchange data values between nodes. The controller
distributes simulation state among nodes by instantiating one or more
partitions of simulation over each node.

As new jobs are submitted to controller, it builds the job graph from their
meta data. The controller uses the job dependencies (before set) and data
dependencies (read/write set) to determine what data values to pass to
computation jobs -- what updates from previous jobs are visible to a job.
Based on the existing distribution of data objects, the
controller picks a target node for executing a job. If the target node does
not have updated data values (e.g. out-dated ghost values), it inserts copy
jobs to exchange data between nodes. A \textit{runtime before set} comprises
of all the computation and copy jobs that must run before a job starts
executing. It ensures that data accesses are race free,
and jobs read correctly updated data. The controller sends a runtime before
set, and data instance identifiers to nodes, when issuing a commmand to execute
a job. A node executes a job only after all the jobs in its runtime before set
complete.

Figure~\ref{fig:architecture} shows one iteration of the simplified
water example. A \verb+FoorLoop+ job
executes on one of the nodes, and spawns new jobs for the next iteration.
The controller issues commands to create data objects $d_1$, $d_2$ and $d'_3$
on node $1$, and sends jobs that operate on the left partition to node $1$.
Similarly, it issues commands to create data objects $d'_2$, $d_3$ and $d_4$,
and sends jobs that operate on the right partition to node $2$.
After the first set of \verb+AdvanceVelocity+ jobs, ghost values on each
node need to be updated from the neighbor.
The controller inserts copy jobs to exchange these.

The controller constantly monitors nodes for their health and performance, and
redistributes data and computations when a node starts straggling or fails. 
It regularly checkpoints a simulation by taking a snapshot of the job graph and
saving simulation state on persistent memory.
Upon failure, it rewinds back to the latest checkpoint and resumes
simulation using the saved simulation data.
The following section discusses load-balancing and fault-tolerance in more
detail.

A design with a centralized controller has two major benefits. First, global
knowledge about cloud resources and their performance helps in detecting
stragglers and failures. Second, control logic for a simulation does not need
to be distributed over multiple nodes -- only the centralized controller needs
metadata for all jobs.
Exchanging job metadata among all nodes to build the job graph at each node
induces a lot of overhead in the cloud, due to large network latencies.

\section{Implementation} \label{sec:implementation}

This section covers three main implementation details required to evaluate
Nimbus abstraction success in running graphical simulations in presence of
stragglers and failures. First, we explain negligible effort in porting current
applications into Nimbus.  Next, the details of providing load balancing and
fault tolerance features are covered. There are a lot of details including
controller optimizations that explaining them is out of the scope of this
paper.

\subsection{Porting Applications} \label{sec:porting}
We have ported water and smoke simulation from PhysBAM library into
the introduced abstraction, by wrapping existing PhysBAM function calls with
Nimbus job abstraction, and adding two loop jobs that spawns the main-loop
and the solver-loop with correct job meta data. There are helper functions
that help specify read/write/before set, and thus the required changes are
small. All in all, water (smoke) simulation required about $2,000$
($1,300$) additional lines of C++ code to be ported compared to the implemented
simulation logic in PhysBAM with over $100,000$ lines of C++ code.

Note that, PhysBAM computations expect to operate over a contiguous data
whereas, in our abstraction data is split into disjoin objects. To Eliminate
any changes in the code base, we implemented a \textit{Translator Layer} that
translates between the disjoint objects and contiguous data back and forth. The
translation happens partially for only the updated objects within the
contiguous data. Explaining the details and intricacies of this layer is out of
the scope of this paper.

\subsection{Load Balancing} \label{sec:lb}

Controller tries to distribute computation work uniformly among all nodes by
adjusting the simulation region each node is responsible for and assigning 
jobs accordingly. It carves out the whole simulation region into contiguous regions and
ties each region to a node. The target node for job execution would be the node
with the region that has the most overlap with the objects in the job's
read/write set. Continuous region assignment eliminates the communication between
nodes.

To achieve load balancing,
controller reduces the size of the region assigned to a node once it
detects the node becomes a straggler. The controller detects stragglers by
periodically retrieving performance report from each node.
A node is treated as a straggler if the ratio of computation time over total
time is over a certain percentage and other nodes are blocking on its ghost
cell data transfer.

\subsection{Fault Tolerance} \label{sec:ft}

Controller periodically creates checkpoints of the simulation state to
rewind back from in case of failures.
Simulation states are made persistent to disk during checkpointing,
and are sharded over different nodes and indexed by a distributed key
value store on top of leveldb\cite{leveldb}. 

The states to be checkpointed includes: a snapshot of the job graph,
all the \textit{parent} jobs that submit other jobs to the controller
(e.g. \verb+For Loop+ jobs in Figure~\ref{fig:abstraction}),
and all data objects that the parent jobs or the jobs they spawned might access.
These saved states are enough for the controller to do a complete rewind back.
Upon failure, controller replaces the current job graph with the saved one,
assigns the saved parent jobs to nodes for execution,
and all data objects that might be possibly accessed are restored.
The restored parent jobs will restart the whole simulation from the checkpoint.

\section{Evaluation} \label{sec:evaluation}

This section evaluates how Nimbus performs in presence of stragglers
and failures. All experiments use a 3D simulation of water pouring into
a half full glass~\cite{glass}. We compare Nimbus performance 
to that of Physbam's MPI-based distributed implementation.
All experiments are run on Amazon
EC2; Nimbus controller runs on a $c3.2xlarge$
instance with $15GB$ of RAM and $8$ hyper-threaded cores, while each compute
node is a $c3.large$ instance with $3.75GB$ of RAM and $2$ hyper-threaded
cores\footnote{Compute-optimized c3 instances use Intel Xeon E5-2680 v2 (Ivy
    Bridge) processors that run at 2.8GHZ.}.  Unless otherwise stated, all
experiments run for $3$ frames and the simulation is $256^3$ grid split
into $16$ partitions and distributed over $8$ computation nodes. 
When there are no stragglers or failures, this simulation takes under
$15$ minutes.

\begin{figure}[t]
    \centering
    \includegraphics[width=3.0in]{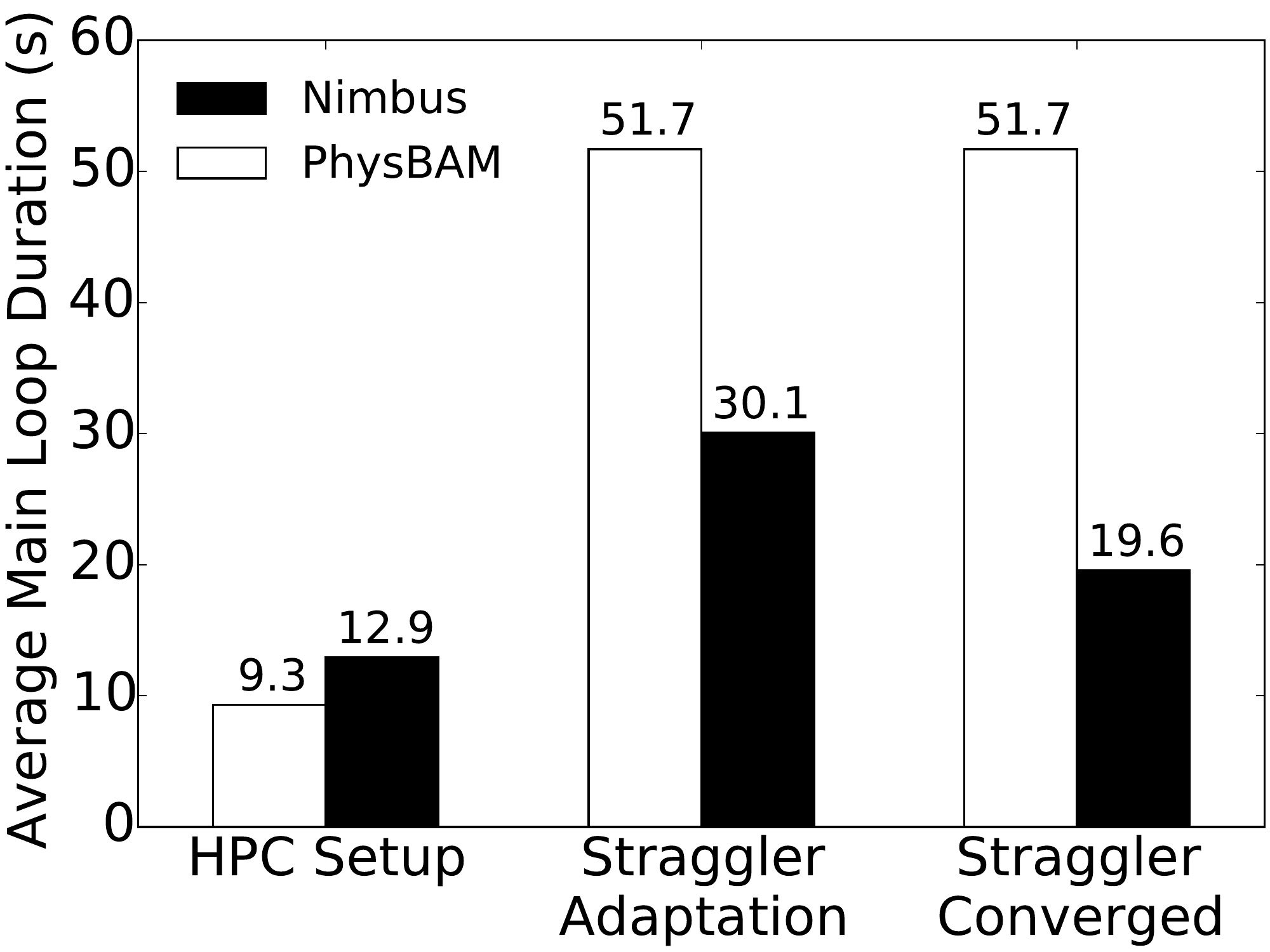}
    \caption{Running a $256^3$ water simulation in HPC setting and cloud
      setting over 8 nodes. Main iteration duration length is measured in
        case of PhysBAM MPI implementation vs. Nimbus. For the cloud settings
        the adaptation and converged periods are separated.}  
\label{fig:lb}
\end{figure}

Figure~\ref{fig:lb} shows how Nimbus and PhysBAM perform with stragglers.
To measure Nimbus' worst case overhead, we first compare its performance
to PhysBAM in an HPC configuration. This is the worst case because it is the
environment PhysBAM was designed for -- there are no stragglers or
failures. Nimbus runs slightly slower in this case. The overhead is primarily
round-trip-times between workers and the controller during the linear solve.
As Table~\ref{tab:sim-scale} shows, for every $61ms$ computation
period there are more than $1500$ data exchange commands issued from controller
to nodes.
To compare performance in presence of stragglers, we evaluate Nimbus and
Physbam when one of nodes starts straggling $5$ minutes into
the simulation. We simulated the straggler by running background processes on
one of the nodes (same method as~\cite{late}). With this straggler and no load
redistribution, the simulation runs $5$ to $6$ times slower~\footnote{As measured
and reported in \cite{mantri}, $10$\% of the outliers are $10X$ slower in the
cloud.}. It takes less than $50$ seconds for the controller to detect and adapt
to the straggler before converging to a balanced load. PhysBAM cannot adapt and
so it goes as slowly as the straggler. However, Nimbus migrates two partitions
at the straggler to other two nodes. This way, the simulation runs
around $1.5X$ slower, as two nodes run $3$ partitions instead of $2$.

\begin{figure}[t]
    \centering
    \includegraphics[width=3.0in]{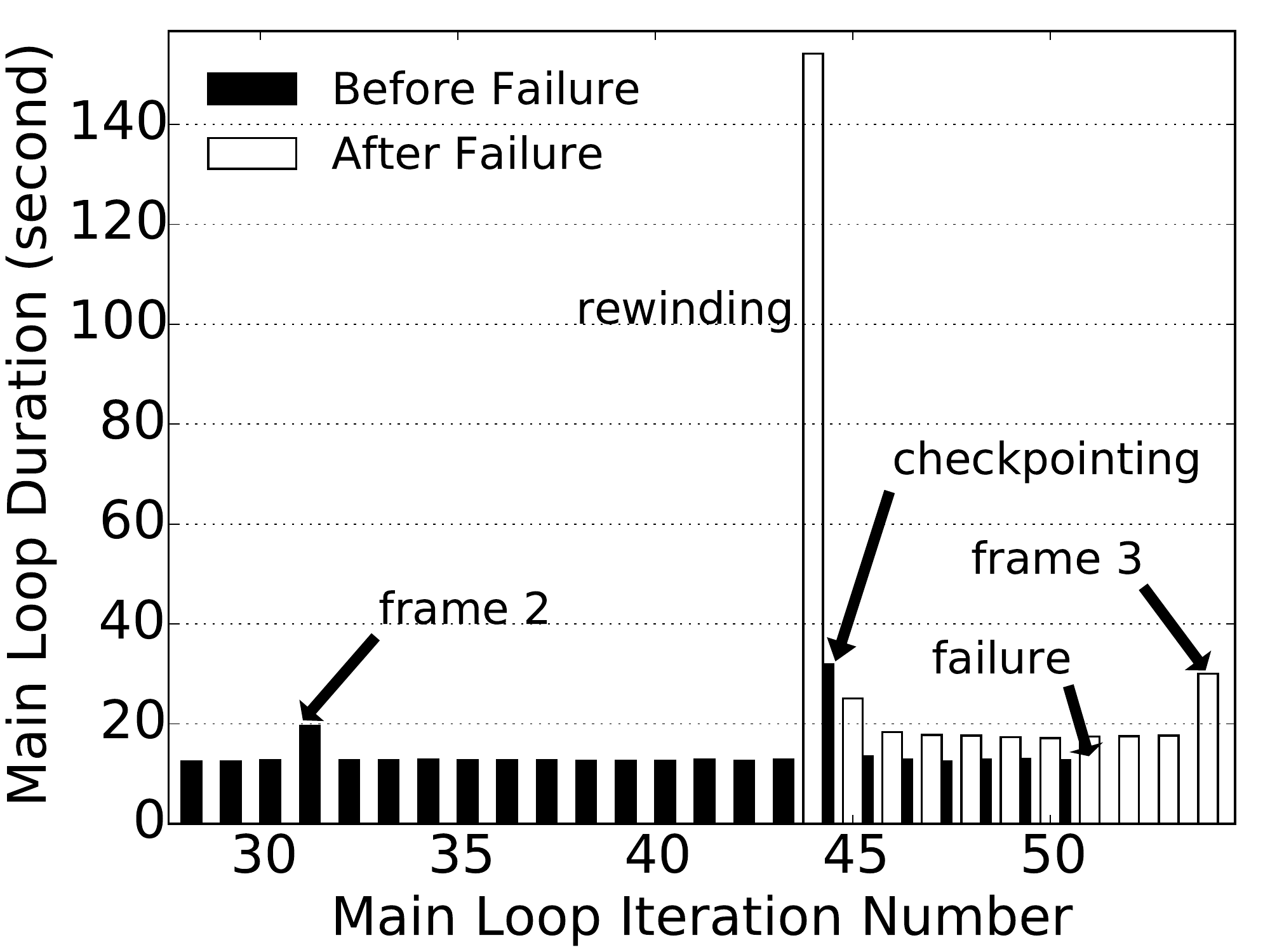}
    \caption{Running a $256^3$ water simulation in presence of failure
      over 8 nodes. Iteration progress is depicted against time.
        Each ripple shows a checkpoint creation. At
        each knee one of the nodes has failed and simulation is reverted back
        by the centralized controller to the last checkpoint.}
\label{fig:ft}
\end{figure}

Last we evaluate Nimbus' fault tolerance mechanisms. In this setup, checkpointing
happens every $10$ minutes, and one of the nodes fails after $11$ minutes into the
simulation.  Figure~\ref{fig:ft} depicts the iteration progress for a time
window.
The controller creates a checkpoint after completion of
$43^{rd}$ iteration, and one of the nodes fails in the middle of computing
$52^{nd}$ iteration. 
Checkpoint creation overhead is less than $18$ seconds.
When the node fails, its in memory state is gone, and
controller rewinds back to the last checkpoint, and recomputes the iterations
from there. The first iteration after rewinding takes around $153$ seconds
which is due to loading aroung $2.4$GB of state from hard disk of remote nodes. Also, iterations take longer after failure because there are less
resources available (same as in the straggler case).

\section{Related Work} \label{sec:related}

Previous work on support for distributing physical simulations, such as
Legion~\cite{legion}, Charm++~\cite{charmpp} and adaptive MPI~\cite{ampi}
have focused on supercomputing and high-performance computing environments.
Legion provides mechanisms to decouple computations from where they run, but
leaves collection and synchronization of runtime information, and actual
load-balancing to applications, and does not provide any fault tolerance.
Charm++ and adaptive MPI load-balance by migrating chare objects and virtual
MPI processes, which do not have any information about geometric locality.
Simulation languages such as Liszt~\cite{liszt} target portability of code, and
use existing mechanisms from the supercomputing domain to parallelize code.
Dandelion~\cite{dandelion} uses a data flow-engine, similar to
Dryad~\cite{dryad}, that is well-suited for parallelism at a coarser granularity.

Existing cloud computing systems such as Map-reduce~\cite{mapreduce} and
Spark~\cite{spark} target highly data parallel computations over key-value
stores.
Systems such as Pregel~\cite{pregel} and Powergraph~\cite{powergraph} target
computations such as scatter and gather over graph data structures.
These computations over key-values and graphs have very different access
patterns compared to graphical simulations over grids.
Nimbus application jobs on the other hand can read and write data at arbitrary
locations in their read and write sets. Application job graphs involve complex
inter-job and data dependencies in Nimbus.

\section{Conclusion and Future Work} \label{sec:future}

Nimbus is a runtime system for running graphical simulation in the cloud.
To utilize cloud resources efficiently, Nimbus addresses problems such as
stragglers and failures by load-balancing and checkpointing.
The key to achieving this is decoupling control flow, computations and data
exchanges. With careful design and optimized data structures, the centralized
controller does not become a bottleneck at common simulation scales.
We have ported a PhysBAM water simulation, an advanced graphical simulation
application, to Nimbus with negligible code changes, and proved that Nimbus can
adapt to cloud performance problems well.

In future, we plan to explore running more partitions per node to have more
flexibility for load balancing, and run even larger simulations. We plan to
examine and address scalability issues when running on a large number of nodes.
The final objective is to be able to run large simulations on hundreds of elastically
provisioned nodes, instead of small and expensive high performance computing clusters.

\bibliography{nimbus}
\bibliographystyle{abbrvnat}

\end{document}